\documentclass[aps,prd,preprintnumbers,nofootinbib,superscriptaddress]{revtex4}
\usepackage{xcolor}
\usepackage{graphicx}
\usepackage{amsfonts}
\usepackage{amssymb}
\usepackage{amsmath}
\usepackage[colorlinks, linkcolor = black, citecolor = black, filecolor = black, urlcolor = black]{hyperref}
\usepackage{multirow}
\usepackage{newtxmath}
\usepackage{xfrac}
\usepackage{makecell}
\usepackage{tabu}
\usepackage{upgreek}
\usepackage{ae}
\usepackage{bm}
\usepackage{slashed}
\usepackage{wasysym}
\usepackage{footnote}
\usepackage[a4paper,textwidth=152mm,top=20mm,textheight=247mm]{geometry}

\begin{document}

\title{Charge-state distributions of highly charged lead ions at relativistic collision energies}

\author{F. M. Kr{\"o}ger}
\email{felix.kroeger@uni-jena.de}
\affiliation{Institut f{\"u}r Optik und Quantenelektronik, Friedrich-Schiller-Universit{\"a}t Jena, Jena, Germany}
\affiliation{Helmholtz-Institut Jena, Jena, Germany}
\affiliation{GSI Helmholtzzentrum f\"ur Schwerionenforschung GmbH, Darmstadt, Germany}
\author{G. Weber}
\affiliation{Helmholtz-Institut Jena, Jena, Germany}
\affiliation{GSI Helmholtzzentrum f\"ur Schwerionenforschung GmbH, Darmstadt, Germany}
\author{S. Hirlaender}
\affiliation{IDA Lab, University of Salzburg, Salzburg, Austria}
\author{R. Alemany-Fernandez}
\affiliation{European Organization for Nuclear Research, Geneva, Switzerland}
\author{M. W. Krasny}
\affiliation{European Organization for Nuclear Research, Geneva, Switzerland}
\affiliation{Laboratoire de physique nucléaire et de hautes energies (LPNHE), University Paris Sorbonne, CNRS‐IN2P3, Paris, France}
\author{Th.~St{\"o}hlker}
\affiliation{Institut f{\"u}r Optik und Quantenelektronik, Friedrich-Schiller-Universit{\"a}t Jena, Jena, Germany}
\affiliation{Helmholtz-Institut Jena, Jena, Germany}
\affiliation{GSI Helmholtzzentrum f\"ur Schwerionenforschung GmbH, Darmstadt, Germany}
\author{\mbox{I. Yu. Tolstikhina}}
\affiliation{P.N. Lebedev Physical Institute, Moscow, Russia}
\author{\mbox{V. P. Shevelko}}
\affiliation{P.N. Lebedev Physical Institute, Moscow, Russia}

\begin{abstract}
Presented is a study of the charge-state evolution of relativistic lead ions passing through a thin aluminum stripper foil. It was motivated by the Gamma Factory project at CERN, where optical laser pulses will be converted into intense gamma-ray beams with energies up to a few hundred MeV via excitation of atomic transitions in few-electron heavy-ions at highly relativistic velocities. In this study all charge-states starting from Pb$^{54+}$ up to bare ions are considered at kinetic projectile energies of 4.2 and 5.9\,GeV/u. To this purpose the BREIT code is employed together with theoretical cross-sections for single-electron loss and capture of the projectile ions. To verify the predicted charge-state evolution, the results are compared to the very few experimental data being available for highly-relativistic lead beams. Reasonable agreement is found, in particular for the yields of Pb$^{80+}$ and Pb$^{81+}$ ions that were recently measured using an aluminum stripper foil located in the transfer beam line between the PS and SPS synchrotron accelerators at CERN. The present study lays the groundwork to optimize the yields of charge states of interest for experiments within the scientific program of the future Gamma Factory project.

\end{abstract}

\maketitle

\section{Introduction}
Charge-changing processes, i.e. loss or capture of electrons, occurring in ion-atom and ion-ion collisions belong to the most basic interactions being present in all types of plasmas as well as at accelerator facilities. When passing through matter, the charge-state distribution of an ion beam approaches with increasing stripper foil thickness the equilibrium distribution, where the fraction of the outgoing charge states are directly related to the individual charge-changing cross-sections but independent from the initial charge state of the ion beams. The equilibrium charge-state distribution, which is commonly characterized by its centroid, also referred to as the equilibrium charge-state, and its width, is determined by the ion species, the velocity of the ion, the target and its phase (solid or gaseous) \cite{SCHIWIETZ2001}. For highly relativistic projectiles, as they are typical for the CERN facility, the cross section for electron loss is much larger as compared to all capture processes, thus the equilibrium charge-state distribution after the passage through a thick stripper target is clearly dominated by the bare charge state. However, as the Gamma Factory project will employ x-ray transitions of few-electron ions for converting optical laser pulses into intense gamma-ray beams, the efficient production of ions in the desired charge states requires the use of thin stripper-foils and consequently dealing with non-equilibrium charge-state distributions of heavy ions \cite{MEYERHOF1987,STOEHLKER1991,SCHEIDENBERGER1998}. To identify the best material and thickness of the stripper target to produce a specific charge state, it is necessary to model the charge-state evolution as a function of the foil thickness.

\begin{figure}[h!]
  \centerline{\includegraphics[width=110mm]{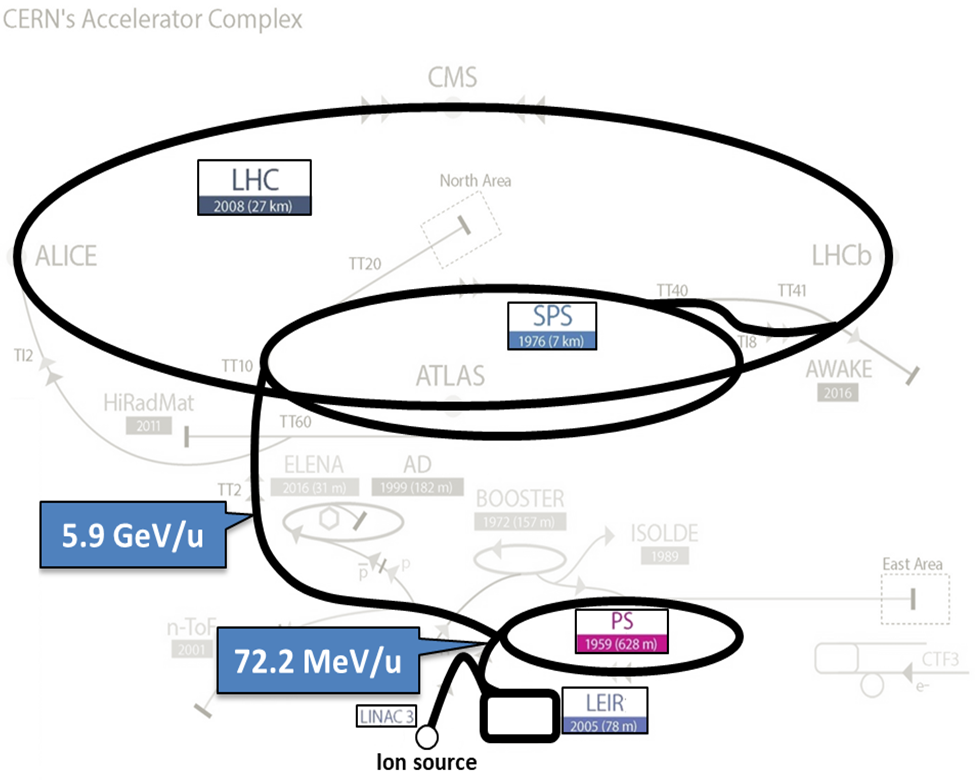}}
  \caption{CERN accelerator complex is shown with all accelerators and rings involved into the production and acceleration of the lead beam. Highlighted are the two stripper foil positions together with the corresponding energies of the two stripping scenario, from which the one at 5.9 GeV/u was chosen, as energy loss and straggling are rather small and of minor importance. (Edited form of \cite{Accelerator2017})}
  \label{fig_CERN}
\end{figure}

In this context recent Gamma Factory beam tests aimed for the efficient production of hydrogen-like (Pb$^{81+}$) and helium-like (Pb$^{80+}$) lead ions. Currently, for this ion species the CERN facility allows two possible stripping scenarios: both are starting with Pb$^{54+}$ ions and then are either using a stripper foil located in the transfer line from the Low Energy Ion Ring (LEIR) to the Proton Synchrotron (PS) at a typical beam energy of 72\,MeV/u or in the TT2 transfer line between the PS and the Super Proton Synchrotron (SPS) at 5.9\,GeV/u (see figure \ref{fig_CERN}). For the Gamma Factory beam tests the high-energy scenario was chosen because at such high energy the energy-loss and -straggling of the ion beam when passing through the stripper foil are negligible and also at the time of experiment the PS control system was not well adapted for the injection of Pb$^{80+}$ or Pb$^{81+}$ beams. Moreover, the efficient production of ions with a K-shell vacancy requires a collision velocity larger than the K-shell orbital velocity. Thus, for achieving a high yield of Pb$^{81+}$ it would have been necessary in any case to employ a second stripping stage at higher energies.

However, the existing setup at TT2 did not allow for the positioning of several stripper foils with various materials and thicknesses to experimentally assess the optimal stripping target. Thus, for a sensible choice of a single foil, it was necessary to model the expected charge-state evolution based on theoretical loss and capture cross-sections for all relevant projectile charge-states. It is important to note, that the built-in cross-sections of commonly available codes for modeling the charge-state evolution of ions passing through matter, i.e. the GLOBAL, ETACHA and CHARGE codes, are not applicable to the domain of highly-relativistic collision-energies and/or are not capable of treating projectile electrons in all the orbitals occupied by the 28 electrons of Pb$^{54+}$. In case of the GLOBAL code, for example, only cross-section computations for energies of up to 2\,GeV/u are possible, while the recent version of the CHARGE code is applicable for highly-relativistic energies, but handles only projectile electrons in the K shell.

Fortunately, the great interest in partially stripped ions at the upcoming Facility for Antiproton and Ion Research (FAIR) in Darmstadt, Germany, has recently motivated the development of the Balance Rate Equations of Ion Transportation (BREIT) code \cite{WINCKLER2017} which can be used to calculate charged-state fractions of relativistic ion beams in target foils (see, e.g. \cite{Shevelko2020,Shevelko20202}). Compared to the aforementioned charge-state distribution codes which have built-in charge-exchange cross-sections covering a specific parameter range, the BREIT code provides a plain solver for the balance rate equations between the various charge states while the underlying cross-sections for capture and loss of projectile electrons need to be supplied by the user. This allows to apply dedicated cross-section predictions tailored to the specific needs of the experiment and to overcome the limitations of the commonly used codes. 

In the following, we compare the results of the BREIT code to the very limited experimental data that are available for highly-relativistic lead beams, namely to the findings of the aforementioned Gamma Factory beam tests as well as to an older measurement of the bare ion (Pb$^{82+}$) yield of aluminum foils of various thicknesses \cite{ARDUINI1997}.

\section{Charge-state distribution modelling and experimental findings}
\begin{figure}[h!]
  \centerline{\includegraphics[width=110mm]{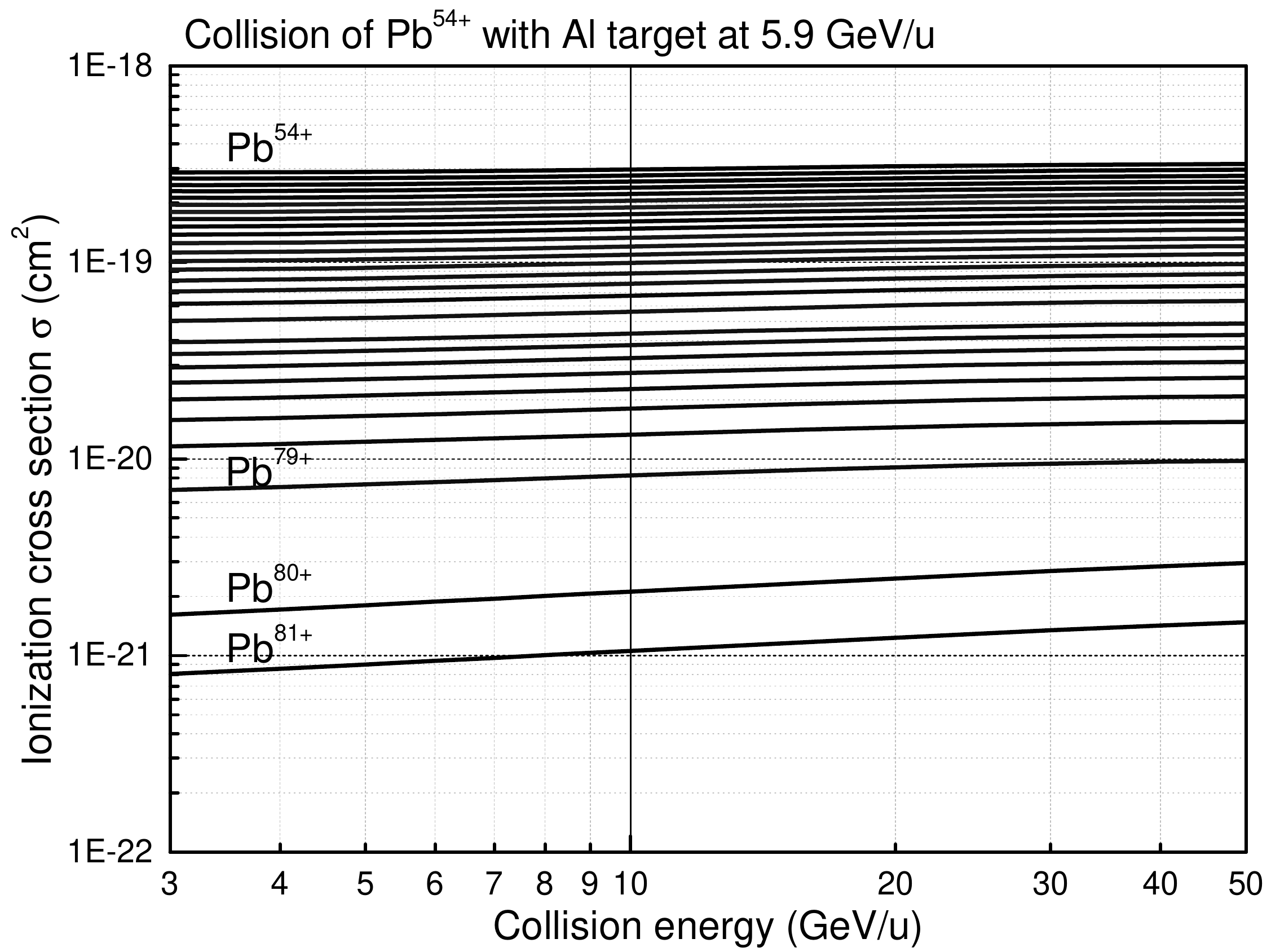}}
  \caption{Ionisation cross sections for different lead ions having different charge states. The cross sections from Pb$^{54+}$ up to Pb$^{79+}$ were calculated using the RICODE-M code, while the cross sections of Pb$^{80+}$ and Pb$^{81+}$ were calculated using the CHARGE code.}
  \label{fig_ionization_cross_sections}
\end{figure}

At highly relativistic collision energies and asymmetric collision conditions (nuclear charge of the projectile much larger compared to the nuclear charge of the target), the cross sections of all relevant electron-capture processes (radiative electron capture, non-radiative electron capture and pair-creation by electron capture) are orders of magnitude smaller than the cross section for the loss of projectile electrons. As a consequence, the prediction of the evolution of the projectile charge-state distribution as a function of penetration depth mainly depends on the knowledge of the underlying electron-loss cross-sections. For the ionization of K-shell electrons, we relied on the prediction of the CHARGE code. Its calculation of loss of single projectile electrons is based on relativistic Born approximation as proposed by Anholt et al. \cite{ANHOLT1987} using tabulated values for proton impact ionization with additional correction factors to take into account higher-order relativistic effects. The predictions are in almost perfect agreement with experimental findings for gold ions colliding with various target materials at an kinetic energy of 11 GeV/u \cite{CLAYTOR1997}. For all higher shells the modified Relativistic Ionization CODE (RICODE‐M) code \cite{TOLSTIKHINA2014} was used to predict cross-sections for single-electron loss of Pb$^{54+}$ to Pb$^{79+}$. These cross sections are presented in figure \ref{fig_ionization_cross_sections}. It is worth noting that in this highly relativistic regime the ionization cross section is only slightly varying as a function of the kinetic beam energy. The mildly increasing trend, that is most pronounced for the K shell, can be attributed of to the enhanced transverse component of the Lorentz-transformed Coulomb fields of the colliding particles. For electron capture, there are three distinct processes that are relevant in the regime of relativistic collision energies, namely radiative electron capture (REC), non-radiative electron capture (NRC) and pair creation where the electron is captured into a bound state of the projectile. REC cross-sections, calculated within a relativistically rigorous manner, were taken from the tabulated values for capture into the K-, L-, and M-shell of Ichihara and Eichler \cite{ICHIHARA2000}. Contributions from REC into higher shells were not taken into account as they are negligible in the relevant energy range. Pair creation was also not taken into account since even with an empty projectile K-shell the measured cross-section is below 1 barn for gold ions impinging on an aluminum target at a kinetic beam energy of 11\,GeV/u \cite{BELKACEM1998}. Non-radiative capture cross-sections were determined based on the relativistic Eikonal approximation as proposed by Eichler \cite{EICHLER1981}. However, the resulting values are several orders of magnitude smaller than the dominant REC contribution and were not taken into account in the BREIT calculations in this work. Also multiple electron-loss and capture processes were neglected as there exists no reliable theory for this energy range. In any case, the contribution of multi-electron processes are expected to play only a minor role for heavy, few electron systems were the few electrons are strongly bound to the projectile. Also two-step processes, where electrons are first excited to a higher lying orbital, from which they are then ionized in a subsequent collision, were neglected as the de-excitation rates for excited projectile states in high-Z ions are rather fast compared to the collision frequency even when taking into account the relativistically enhanced lifetimes.

In order to find the optimal stripper-foil material and thickness, calculations based on the BREIT code were performed for different target foil materials (see \cite{KROEGER2020} for more details). Aluminum was chosen as stripping material for the Gamma Factory beam tests and the appropriate thickness of the foil was determined according to the maximum yields of the charge states of interest (Pb$^{80+}$ and Pb$^{81+}$). As a result, a foil with a thickness of 150\,$\mu m$ was chosen at a tilt angle of 45$^{\circ}$. The effective foil thickness amounts to 212\,$\mu m$. After their successful production (stripping) and injection into the SPS, the Pb$^{80+}$ beam was accelerated to 270$\cdot$Q\,GeV/c, and the Pb$^{81+}$ beam to 450$\cdot$Q\,GeV/c which corresponds to the LHC injection momentum. Thus, the experiment allowed both to test beam lifetimes of stored few-electron Pb ions and to study the production efficiency of Pb$^{81+}$ and Pb$^{80+}$ at relativistic velocities.\\

\begin{figure}[h!]
  \centerline{\includegraphics[width=110mm]{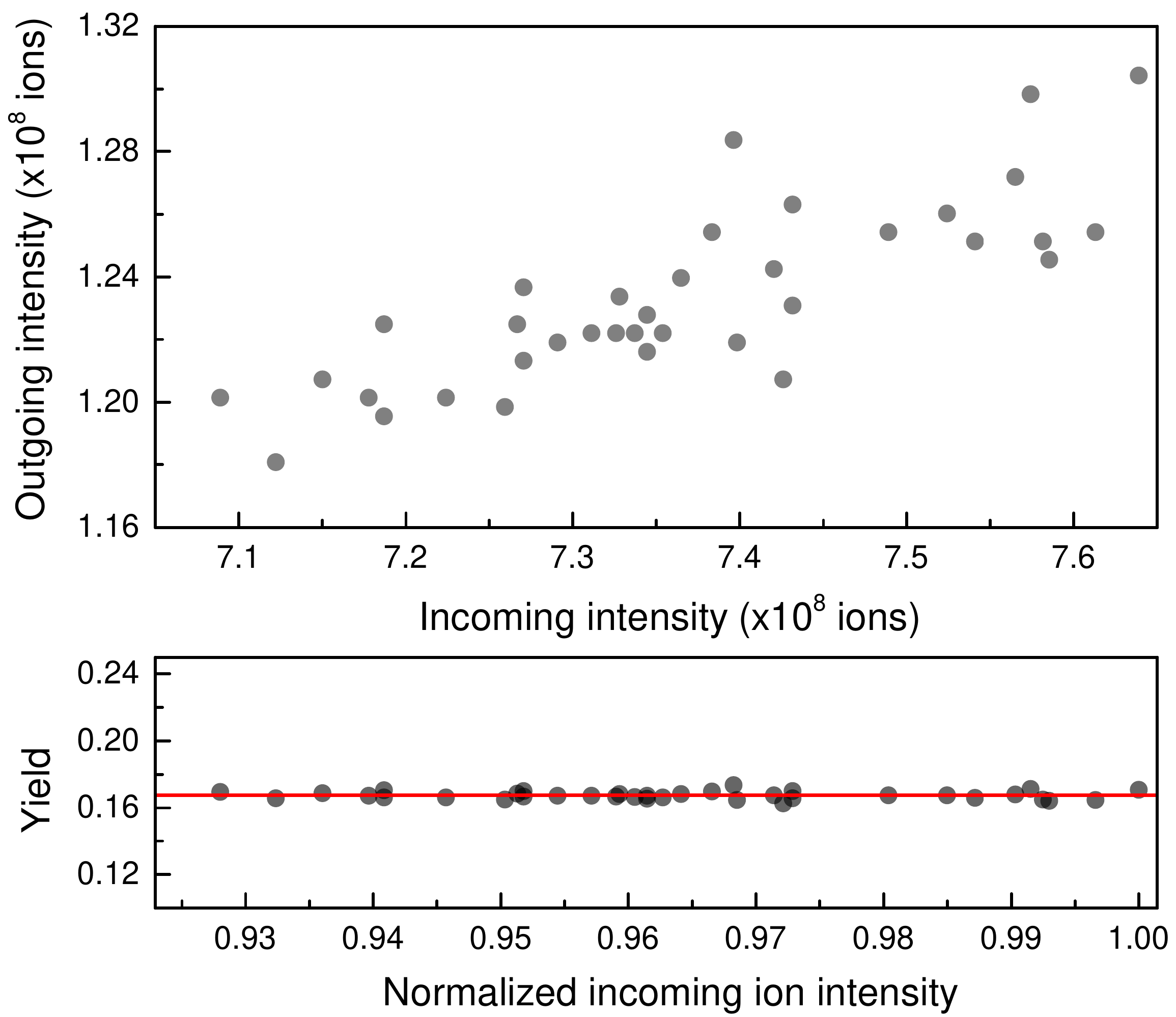}}
  \caption{Upper picture: outgoing Pb$^{80+}$ beam intensity plotted against incoming Pb$^{54+}$ beam intensity. Lower picture: Yield of Pb$^{80+}$ ions plotted against the normalized incoming ion intensity. The mean value of the yield is marked by a red line.}
  \label{fig_Pb80+}
\end{figure}

\begin{figure}[h!]
  \centerline{\includegraphics[width=110mm]{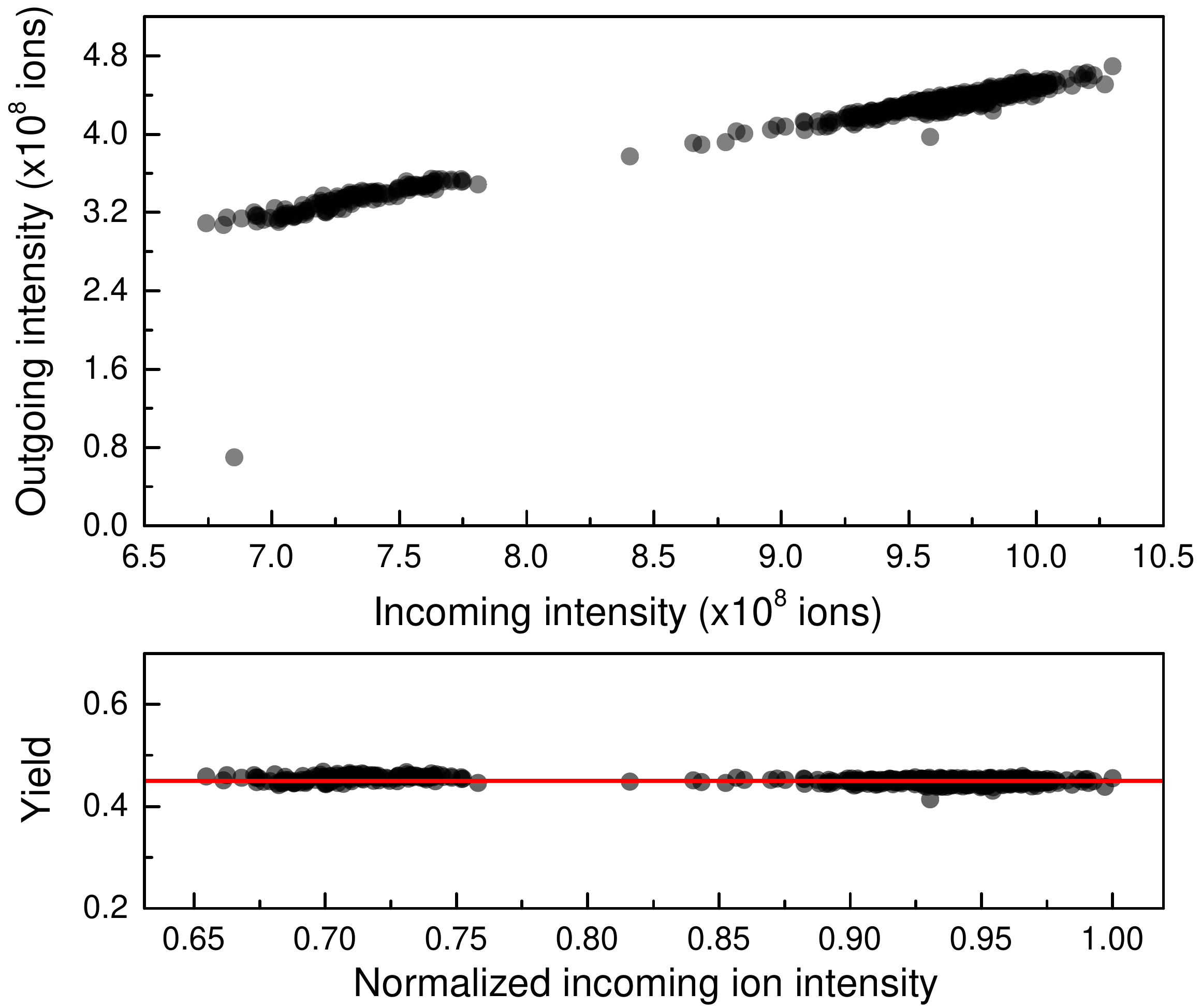}}
  \caption{Upper picture: outgoing Pb$^{81+}$ beam intensity plotted against incoming Pb$^{54+}$ beam intensity. The escaped data point at low intensity seems to indicate a defective exception and was therefore excluded from the data analysis. Lower picture: Yield of Pb$^{81+}$ ions plotted against the normalized incoming ion intensity. The mean value of the yield is marked by a red line.}
  \label{fig_Pb81+_long}
\end{figure}

The ion beam intensity along the accelerator chain was measured by the beam current transformers installed at the PS and SPS rings as well as in the TT2 transfer line (before and after stripping). The beam intensities of Pb$^{80+}$ and Pb$^{81+}$ after the passage through the stripper foil are shown in the upper pictures of figures \ref{fig_Pb80+} and \ref{fig_Pb81+_long} relative to the intensity of the incoming Pb$^{54+}$ beam. The lower pictures show the outgoing ion beam intensity normalized to the incoming intensity, i.e. the production yield for the charge state of interest, as well as the mean value of the yield as a red line. The data set presented in figure \ref{fig_Pb81+_long} covers two different intensity ranges allows to estimate a systematic measurement error. The mean yields of both subsets deviates by about 0.7\% from each other, while the mean standard deviation of the data set in figure \ref{fig_Pb80+} as well as the two in figure \ref{fig_Pb81+_long} corresponds to roughly 0.4\%. Thus, one can conclude that the measurement uncertainty of the charge state yields that are caused by shot-to-shot deviations of the accelerator facility and by different ion beam intensities is of the order of 1\,\% and below. In addition there is a systematic uncertainty stemming from the calibration of the beam current transformers which is estimated to be on the same level.

\begin{figure}[h!]
  \centerline{\includegraphics[width=110mm]{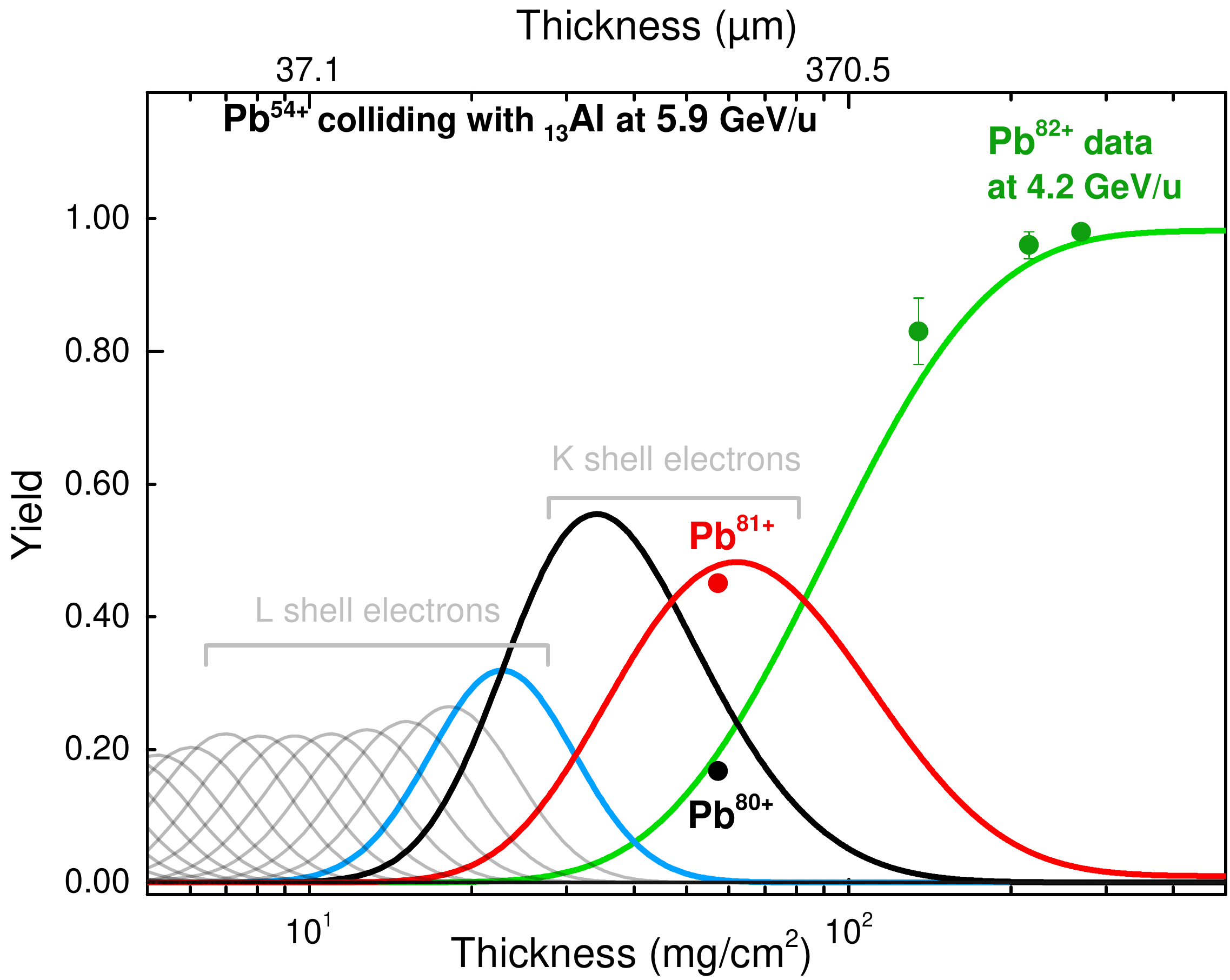}}
  \caption{Comparison of the BREIT code calculation results with experimental data for Pb$^{80+}$, Pb$^{81+}$ from the Gamma Factory beam tests at 5.9\,GeV/u, and for Pb$^{82}$+ from \cite{ARDUINI1997} at 4.2\,GeV/u . If the systematic uncertainty bars of the experimental data points can not be seen, it is because they are comparably small to the data points, see text for more details. In addition the blue line highlights the prediction result for the production of Pb$ ^{79+} $ ions which will be subject of an upcoming Gamma Factory Proof-of-Principle experiment.}
  \label{fig_charge_state_distribution}
\end{figure}

In figure \ref{fig_charge_state_distribution} the experimental findings are compared to the results obtained with the BREIT code. As can be seen, there is a reasonable agreement between the predictions and the experimental data. This finding proofs, that the applied approach in selecting the appropriate cross-sections in combination with the BREIT code enables the reliable prediction of non-equilibrium charge-state distributions even at relativistic energies as high as 5.9\,GeV/u. While the deviation of the measured yield of Pb$^{80+}$ seems comparably large compared to the one of Pb$^{81+}$, it should be noted that data of partially stripped ions in this highly relativistic energy region are very scarce, restraining the possibility to test and benchmark codes and theories. By this measure, the nearly 50\% deviation is still good, especially as it is directly connected to the uncertainty estimation of the used ionization cross sections which is assumed to be about the same.

Of course, one may note that highly interesting information about atomic charge-exchange processes for partially stripped ions at high relativistic energies have already been obtained in previous experiments at the Alternating Gradient Synchrotron (AGS) at Brookhaven National Laboratory as well as at the SPS at CERN. However, this was typically done by using initially bare ions penetrating through thin target foils obtaining typical yields for H-like heavy ions of below 1\% (e.g. Au$^{78+}$). To the best of our knowledge, in the present work, the experimental study of partially stripping of heavy ions at highly relativistic energies is reported for the very first time. In fact, this allowed to optimize for the yield of not fully ionized ions such as Pb$^{81+}$ with a sizeable fraction of close to 50\% of the initial Pb$^{54+}$ beam. This provides unique access to atomic processes on ultra-fast atomic time-scales and enables to benchmark corresponding theoretical models. For this reason we also show in figure \ref{fig_charge_state_distribution} experimental data of Pb$^{82+}$ at 4.2\,GeV/u \cite{ARDUINI1997} in comparison with the results of corresponding  calculations based on the BREIT code. Again, the computed results are in good agreement with the experimental findings. Nevertheless, one may argue that there is a systematic shift between experimental data and the theoretical results. As aforementioned, the charge-state distribution in this high-energy range is determined by the ionization cross-sections. This allows the assumption, that either the employed ionization cross-sections are systematically predicted (slightly) too small, or that this is caused by other ionizing effects that could not be taken into account since appropriated theoretical models are not available. As such one has to mention, two-step ionization (electron excitation and ionization in subsequent collisions) which is in general not negligible in this energy range any more for electrons in the L and M shell. Also saturation effects for the cross-sections due to the strongly increasing transverse electric field component with increasing $\gamma$ might already start to affect the ionization cross-sections (see \cite{SOERENSEN1998}). But as the single-electron ionization is still estimated to be by far the dominant process, and since the uncertainty for the L-shell ionization (and the ionization of electrons in higher shells) is estimated to be of the order of 50\%, we assume that slight deviations are mostly arising from the uncertainty of the ionization cross-sections for the L- and higher shells. Finally, we note that the uncertainty of experimental parameters such as the stripper foil thickness are difficult to estimate but are expected to be lower than 10\%.\\

It should be mentioned that an upcoming Gamma Factory Proof-of-Principle (PoP) experiment is planned, that will in particular focus on the production of Pb$^{79+}$ (Li-like lead) in this high energy range. As for the efficient production of this charge state a stripper foil thickness is required which is much smaller compared to our current study (see the blue line in figure \ref{fig_charge_state_distribution}), the uncertainty related to our theoretical approach is expected to be larger as the error tolerance is smaller. However, while for the Gamma Factory beam tests only one single stripper foil was used and tilted in a fixed position, for the PoP experiment a remote-controlled stripper target station will be available which will allow to change, during the beam time, between different foils. This approach will enable an experimental optimization process to determine the most efficient stripping foil for the desired charge-state.

\section{Conclusion}
In summary, in a preparatory study for the Gamma Factory project it was possible to prove the feasibility to produce both Pb$^{80+}$ as well as Pb$^{81+}$ starting with Pb$^{54+}$ at highly relativistic beam velocities at the CERN accelerator facility. On basis of the BREIT code used along with the described relativistic charge-exchange cross-sections, we were able to predict the measured non-equilibrium charge-state distribution very well. This reinforces the importance of this Gamma Factory beam tests data, as it allows to benchmark the used calculation model for further calculations in this newly available energy region for partially stripped lead ions. Nevertheless, besides the ionization cross-sections also the inclusion of further ionizing charge-exchange effects, like multi-electron ionization, would probably diminish the found deviation. While their influence is estimated to be comparably small to the single-electron ionization, they still are not negligible any more. For the moment though, there are no corresponding theories available yet for the given experimental parameters.\\

For the near future, a Gamma Factory Proof-of-Principle experiment is planned which will in particular focus on the production of Pb$ ^{79+} $ in this high energy range. In this context we have also shown prediction results for the preparation of the necessary ion beam stripping. While the uncertainty related to our theoretical approach is expected to be larger as the error tolerance is smaller, there will be on the other hand, a new stripper target station available that will enable an experimental optimization process to determine the most efficient stripping foil for the desired charge-state. Newly measured data will then allow further benchmarking of the used calculation model.

\medskip
\textbf{Acknowledgements} \par
\textit{The author F. M. Kröger acknowledges funding from BMBF via Verbundforschung 05P2018.}

\medskip

\bibliographystyle{ieeetr}
\bibliography{references}

\newpage

\end{document}